\documentstyle[aps,amssymb]{revtex}

\begin{document}
\title{Decoherence in non integrable systems.}
\author{M. Castagnino}
\address{Institutos de F\'{i}sica Rosario y de Astronom\'{i}a y F\'{i}sica del\\
Espacio.\\
Casilla de Correos 67, Sucursal 28, 1628 Buenos Aires, Argentina \\
e-mail: mariocastagnino@citynet.net.ar}
\maketitle

\begin{abstract}
Self-induced decoherence formalism and the corresponding classical limit are
extended from quantum integrable systems to non-integrable ones.
\end{abstract}

\section{Introduction.}

Decoherence was initially considered to be produced by {\it destructive
interference}\cite{vK}. Later the strategy changed and decoherence was
explained as caused by the interaction with an environment\cite{Zurek}, but
this approach is not conclusive because:

i.- The environment cannot always be defined, e. g. in closed system like
the universe.

ii.-There is not a clear definition of the ''cut'' between the proper system
and its environment.

iii.- The definition of the {\it pointer basis }is not simple.{\it \ }

So we need a new and complete theory: {\it The self-induced approach}\cite
{SI}{\it ,} based in a new version of destructive interference, which will
be explained in this talk in its version for non-integrable systems. The
essential idea is that this interference is embodied in Riemann-Lebesgue
theorem where it is proved that if $f(\nu )\epsilon {\Bbb L}_{1}$ then 
\[
\lim_{t\rightarrow \infty }\int_{-a}^{a}f(\nu )e^{-i\frac{\nu t}{\hbar }%
}dt=0 
\]
If we use this formula in the case when $\nu =\omega -\omega ^{\prime }$,
where $\omega ,\omega ^{\prime }$ are the indices of the density operator $%
\widehat{\rho }$, in such a way that $\nu =0$ corresponds to the diagonal,
we obtain a {\it catastrophe, }since all {\it diagonal} and {\it not diagonal%
} terms would disappear. But, if $f(\nu )=A\delta (\nu )+f_{1}(\nu ),$ where
now $f_{1}(\nu )\epsilon {\Bbb L}_{1}$, we have 
\[
\lim_{t\rightarrow \infty }\int_{-a}^{a}f(\nu )e^{-i\frac{\nu t}{\hbar }%
}dt=A 
\]
and the diagonal terms $\nu =0$ remain while the off-diagonal ones vanish.
This is the trick we will use below.

\section{Weyl-Wigner-Moyal mapping.}

Let ${\cal M=M}_{2(N+1)}\equiv {\Bbb R}^{2(N+1)}$ be the phase space.$_{%
\text{ }}$The functions over ${\cal M}$ will be called $f(\phi ),$ where $%
\phi $ symbolizes the coordinates of ${\cal M}$%
\[
\phi ^{a}=(q^{1},...,q^{N+1},p_{q}^{1},...,p_{q}^{N+1}) 
\]
Then the Wigner transform reads

\[
symb\widehat{f}\circeq f(\phi )=\int \langle q+\Delta |\widehat{f}|q-\Delta
\rangle e^{i\frac{p\Delta }{\hbar }}d^{N+1}\Delta 
\]
where $\widehat{f}\epsilon \widehat{{\cal A}}$ and $f(\phi )\epsilon {\cal A}
$ where $\widehat{{\cal A}}$ is the quantum algebra and the classical one is 
${\cal A}$. We can also introduce the star product 
\[
symb(\widehat{f}\widehat{g})=symb\widehat{f}*symb\widehat{g}=(f*g)(\phi
),\qquad (f*g)(\phi )=f(\phi )\exp \left( -\frac{i\hbar }{2}\overleftarrow{%
\partial }_{a}\omega ^{ab}\overrightarrow{\partial }_{b}\right) g(\phi ) 
\]
and the {\it Moyal bracket, }which is the symbol corresponding to the
commutator 
\[
\{f,g\}_{mb}=\frac{1}{i\hbar }(f*g-g*f)=symb\left( \frac{1}{i\hbar }%
[f,g]\right) 
\]
so we have 
\begin{equation}
(f*g)(\phi )=f(\phi )g(\phi )+0(\hbar ),\qquad
\{f,g\}_{mb}=\{f,g\}_{pb}+0(\hbar ^{2})  \label{L.2.12}
\end{equation}
To obtain the inverse $symb^{-1}$ we will use the {\it symmetrical} or {\it %
Weyl} ordering prescription, namely 
\[
symb^{-1}[q^{i}(\phi )p^{j}(\phi )]=\frac{1}{2}\left( \widehat{q}^{i}%
\widehat{p}^{j}+\widehat{p}^{j}\widehat{q}^{i}\right) 
\]
Then we have an isomorphism between the quantum algebra $\widehat{{\cal A}}$
and the classical one ${\cal A}$%
\[
symb^{-1}:{\cal A\rightarrow }\widehat{{\cal A}},\quad symb:\widehat{{\cal A}%
}{\cal \rightarrow A} 
\]
The mapping so defined is the {\it Weyl-Wigner-Moyal symbol}.

For the state we have 
\[
\rho (\phi )=symb\widehat{\rho }=(2\pi \hbar )^{-N-1}symb_{\text{(for
operators)}}\widehat{\rho } 
\]
and it turns out that 
\begin{equation}
(\widehat{\rho }|\widehat{O})=(symb\widehat{\rho }|symb\widehat{O})=\int
d\phi ^{2(N+1)}\rho (\phi )O(\phi )  \label{F}
\end{equation}
Namely the definition $\widehat{\rho }\epsilon \widehat{\text{ }{\cal A}%
^{\prime }},$ as afunctional on $\widehat{{\cal A}},$ is equal to the
definition $symb\rho \epsilon $ ${\cal A}^{\prime },$ as afunctional on $%
{\cal A}.$

\section{Decoherence in non integrable systems.}

\subsection{Local CSCO.}

a.- When our quantum system is endowed with a CSCO of $N+1$ observables,
containing $\widehat{H,}$ the underlying classical system is {\it integrable}%
. In fact, let $N+1-$CSCO be $\{\widehat{H,}\widehat{O}_{1},...,\widehat{O}$ 
$_{N}\}$ the Moyal brackets of these quantities are 
\[
\{O_{I}(\phi ),O_{J}(\phi )\}_{mb}=symb\left( \frac{1}{i\hbar }[\widehat{O}%
_{I},\widehat{O}_{J}]\right) =0 
\]
where $I,$ $J,...=0,1,...,N$ and $\widehat{H}=\widehat{O}_{0}.$ Then when $%
\hbar \rightarrow 0$ from eq. (\ref{L.2.12}) we know that 
\begin{equation}
\{O_{I}(\phi ),O_{J}(\phi )\}_{pb}=0  \label{A}
\end{equation}
then as $H(\phi )=O_{0}(\phi )$ the set $\{O_{I}(\phi )\}$ is a complete set
of $N+1$ constants of the motion in involution, globally defined over all $%
{\cal M}$, and therefore the system is integrable. q. e. d.

b.- If this is not the case $N+1$ constants of the motion in involution $%
\{H,O_{1},...,O$ $_{N}\}$ {\it always exist locally}, as can be shown
integrating the system of equations (\ref{A}). Then, if $\phi _{i}\epsilon 
{\cal M}$ there is {\it maximal domain of integration }${\cal D}_{\phi _{i}}$%
{\it \ around} $\phi _{i}\epsilon {\cal M}$ where these constants are
defined. In this case the system in {\it non-integrable}. Moreover we can
repeat the procedure with the system 
\begin{equation}
\{O_{I}(\phi ),O_{J}(\phi )\}_{mb}=0
\end{equation}
Then we can extend the definition of the constant $\{H,O_{1},...,O$ $_{N}\},$
defined in each ${\cal D}_{\phi _{i}},$ outside ${\cal D}_{\phi _{i}}$ as
null functions. Their Weyl transforms $\{\widehat{H,}\widehat{O}_{1},...,%
\widehat{O}$ $_{N}\}$ can be considered as a local $N+1$-CSCOs related each
one with a domain ${\cal D}_{\phi _{i}}$ that we will call $\{\widehat{H,}%
\widehat{O}_{1\phi _{i}},...,\widehat{O}$ $_{N\phi _{i}}\}$ (we consider
that $\widehat{H}$ is always globally defined).

c.-We also can define an {\it ad hoc positive partition of the identity } 
\[
1=I(\phi )=\sum_{i}I_{\phi _{i}}(\phi ) 
\]
where $I_{\phi _{i}}(\phi )$ is the {\it characteristic} function or {\it %
index} function, i. e.: 
\[
I_{\phi _{i}}(\phi )=\left\{ 
\begin{array}{l}
1\text{ if }\phi \epsilon D_{\phi _{i}} \\ 
0\text{ if }\phi \notin D_{\phi _{i}}
\end{array}
\right. 
\]
where the domains $D_{\phi _{i}}\subset {\cal D}_{\phi _{i}}$ $D_{\phi _{i}}$
$\cap $ $D_{\phi _{j}}=\emptyset $. Then $\sum_{i}I_{\phi _{i}}(\phi )=1.$
Then we can define $A_{\phi _{i}}(\phi )=A(\phi )I_{\phi _{i}}(\phi )$ and 
\[
A(\phi )=\sum_{i}A_{\phi _{i}}(\phi ) 
\]
and using $symb^{-1}$ 
\[
\widehat{A}=\sum_{i}\widehat{A}_{\phi _{i}} 
\]
We can further decompose 
\begin{equation}
\widehat{A}_{\phi _{i}}=\sum_{j}A_{j\phi _{i}}|j\rangle _{\phi _{i}}\langle
j|_{\phi _{i}}  \label{1}
\end{equation}
where the $|j\rangle _{\phi _{i}}$ are the corresponding eigenvectors of the
local $N+1-$CSCO of $D_{\phi _{i}}$ $\subset {\cal D}_{\phi _{i}}$ where a
local $N+1$-CSCO is defined.$.$ So 
\[
\widehat{A}=\sum_{ij}A_{j\phi _{i}}|j\rangle _{\phi _{i}}\langle j|_{\phi
_{i}} 
\]
all over ${\cal M.}$ It can be proved that for $i\neq k$ it is 
\[
\langle j|_{\phi _{i}}|j\rangle _{\phi _{k}}=0 
\]
so the last decomposition is orthonormal, thus decomposition (\ref{1})
generalizes the usual eigen-decomposition of integrable system to the
non-integrable case. We will use this decomposition below.

\subsection{Decoherence in the energy.}

a.- Let us define {\it in each }$D_{\phi _{i}}$ a local $N+1-$CSCO \{$%
\widehat{H}$ ,$\widehat{O_{\phi _{i}}}\}$ (as we have said we consider that $%
\widehat{H}$ is always globally defined) as 
\[
\widehat{H}=\int_{0}^{\infty }\omega \sum_{im}|\omega ,m\rangle _{\phi
_{i}}\langle \omega ,m|_{\phi _{i}}d\omega ,\qquad \widehat{O_{\phi _{i}I}}%
=\int_{0}^{\infty }\sum_{m}O_{m_{I\phi _{i}}}|\omega ,m\rangle _{\phi
_{i}}\langle \omega ,m|_{\phi _{i}}d\omega 
\]
where we have used decomposition (\ref{1}). The energy spectrum is $0\leq
\omega <\infty $ and $m_{I\phi _{i}}=\{m_{1\phi _{i}},...,m_{N\phi
_{i}}\},m_{I\phi _{i}}\epsilon {\Bbb N}$. Therefore 
\[
\quad \widehat{H}|\omega ,m\rangle _{\phi _{i}}=\omega |\omega ,m\rangle
_{\phi _{i}},\qquad \widehat{O_{\phi _{i}I}}|\omega ,m\rangle _{\phi
_{i}}=O_{m_{I\phi _{i}}}|\omega ,m\rangle _{\phi _{i}}. 
\]
where, from the orthomormality of the eigenvector and eq. (\ref{1}), we have 
\[
\langle \omega ,m|_{\phi _{i}}|\omega ^{\prime },m^{\prime }\rangle _{\phi
_{j}}=\delta (\omega -\omega ^{\prime })\delta _{mm^{\prime }}\delta _{ij} 
\]

b.- A generic observable, in the orthonormal basis just defined, reads: 
\[
\widehat{O}=\sum_{imm^{\prime }}\int_{0}^{\infty }\int_{0}^{\infty }d\omega
d\omega ^{\prime }\widetilde{O}(\omega ,\omega ^{\prime })_{\phi
_{i}mm^{\prime }}|\omega ,m\rangle _{\phi _{i}}\langle \omega ^{\prime
},m^{\prime }|_{\phi _{i}} 
\]
where $\widetilde{O}(\omega ,\omega ^{\prime })_{\phi _{i}mm^{\prime }}$ is
a generic {\it kernel} or {\it distribution} in $\omega ,$ $\omega ^{\prime
}.$ As explained in the introduction, the simplest choice to solve our
problem is the van Hove choice\cite{vH}. 
\begin{equation}
\widetilde{O}(\omega ,\omega ^{\prime })_{\phi _{i}mm^{\prime }}=O(\omega
)_{\phi _{i}mm^{\prime }}\delta (\omega -\omega ^{\prime })+O(\omega ,\omega
^{\prime })_{\phi _{i}mm^{\prime }}  \label{2}
\end{equation}
where we have a {\it singular} and a {\it regular} term, so called because
the first one contains a Dirac delta and in the second one the $O(\omega
,\omega ^{\prime })_{\phi _{i}mm^{\prime }}$ are ordinary functions of the
real variables $\omega $ and $\omega ^{\prime }$. As we will see these two
parts appear in every formulae below. So our operators belong to an algebra $%
\widehat{{\cal A}}$ and they read 
\[
\widehat{O}=\sum_{imm^{\prime }}\int_{0}^{\infty }d\omega O(\omega )_{\phi
_{i}mm^{\prime }}|\omega ,m\rangle _{\phi _{i}}\langle \omega ,m^{\prime
}|_{\phi _{i}}+\sum_{imm^{\prime }}\int_{0}^{\infty }\int_{0}^{\infty
}d\omega d\omega ^{\prime }O(\omega ,\omega ^{\prime })_{\phi _{i}mm^{\prime
}}|\omega ,m\rangle _{\phi _{i}}\langle \omega ^{\prime },m^{\prime }|_{\phi
_{i}} 
\]
The {\it observables} are the self adjoint $O^{\dagger }=O$ operators$.$
These observables belong to a space $\widehat{{\cal O}}\subset $ $\widehat{%
{\cal A}\text{ }}$ . This space has the {\it basis} \{$|\omega ,m,m^{\prime
})_{\phi _{i}}$, $|\omega ,\omega ^{\prime },m,m^{\prime })_{\phi _{i}}\}$
defined as: 
\[
\quad |\omega ,m,m^{\prime })_{\phi _{i}}\doteq |\omega ,m\rangle _{\phi
_{i}}\langle \omega ,m^{\prime }|_{\phi _{i}},\qquad |\omega ,\omega
^{\prime },m,m^{\prime })_{\phi _{i}}\doteq |\omega ,m\rangle _{\phi
_{i}}\langle \omega ^{\prime },m^{\prime }|_{\phi _{i}} 
\]
c.- Let us define the quantum states $\widehat{\rho }$ $\in \widehat{{\cal S}%
}{\cal \subset }\widehat{{\cal O}}^{^{\prime }}$, where $\widehat{{\cal S}}$
is a convex set. The basis of $\widehat{{\cal O}}^{\prime }$ is \{$(\omega
,mm^{\prime }|_{\phi _{i}}$, $(\omega \omega ^{\prime },mm^{\prime }|_{\phi
_{i}}\}$ and its vectors are defined as functionals by the equations: 
\[
\quad (\omega ,m,m^{\prime }|_{\phi _{i}}|\eta ,n,n^{\prime })_{\phi
_{j}}=\delta (\omega -\eta )\delta _{mn}\delta _{m^{\prime }n^{\prime
}}\delta _{ij},\qquad (\omega ,\omega ^{\prime },m,m^{\prime }|_{\phi
_{i}}|\eta ,\eta ^{\prime },n,n^{\prime })_{\phi _{j}}=\delta (\omega -\eta
)\delta (\omega ^{\prime }-\eta ^{\prime })\delta _{mn}\delta _{m^{\prime
}n^{\prime }}\delta _{ij}. 
\]
and all others $(.|.)$ are zero. Then, a generic quantum state reads: 
\[
\widehat{\rho }=\sum_{imm^{\prime }}\int_{0}^{\infty }d\omega \overline{\rho
(\omega )}_{\phi _{i}mm^{\prime }}(\omega ,mm^{\prime }|_{\phi
_{i}}+\sum_{imm^{\prime }}\int_{0}^{\infty }d\omega \int_{0}^{\infty
}d\omega ^{\prime }\overline{\rho (\omega ,\omega ^{\prime })}_{\phi
_{i}mm^{\prime }}(\omega \omega ^{\prime },mm^{\prime }|_{\phi _{i}} 
\]
We require that: 
\begin{equation}
\quad \overline{\rho (\omega ,\omega ^{\prime })}_{\phi _{i}mm^{\prime
}}=\rho (\omega ^{\prime },\omega )_{\phi _{i}m^{\prime }m},\text{ }\rho
(\omega ,\omega )_{\phi _{i}mm}\geq 0,\text{ }(\widehat{\rho }|\widehat{I}%
)=\sum_{im}\int_{0}^{\infty }d\omega \rho (\omega )_{\phi _{i}}=1,  \label{Z}
\end{equation}
where $\widehat{I}=\int_{0}^{\infty }d\omega \sum_{im}|\omega ,m\rangle
_{\phi _{i}}\langle \omega ,m|_{\phi _{i}}$ is the identity operator. Then,
in fact, $\widehat{\rho }$ $\in \widehat{{\cal S}}$, where $\widehat{{\cal S}%
}$ is a convex set, and we have 
\begin{equation}
\langle \widehat{O}\rangle _{\widehat{\rho }(t)}=(\widehat{\rho }(t)|%
\widehat{O})=\sum_{imm^{\prime }}\int_{0}^{\infty }d\omega \overline{\rho
(\omega )}_{\phi _{i}mm^{\prime }}O(\omega )_{\phi _{i}mm^{\prime
}}+\sum_{imm^{\prime }}\int_{0}^{\infty }d\omega \int_{0}^{\infty }d\omega
^{\prime }\overline{\rho (\omega ,\omega ^{\prime })}_{\phi _{i}mm^{\prime
}}e^{i(\omega -\omega ^{\prime })t/\hbar }O(\omega ,\omega ^{\prime })_{\phi
_{i}mm^{\prime }}  \label{W}
\end{equation}
If we now take the limit $t\rightarrow \infty $ and use the Riemann-Lebesgue
theorem, being $O(\omega ,\omega ^{\prime })$ and $\overline{\rho (\omega
,\omega ^{\prime })}_{\phi _{i}mm^{\prime }}$ regular (namely $^{\prime }%
\overline{\rho (\omega ,\omega ^{\prime })}_{\phi _{i}mm^{\prime }}O(\omega
,\omega ^{\prime })\epsilon {\Bbb L}_{1}$ in the variable $\nu =\omega
-\omega ^{\prime }),$ we arrive to 
\[
\lim_{t\rightarrow \infty }\langle \widehat{O}\rangle _{\widehat{\rho }(t)}=(%
\widehat{\rho }_{*}|\widehat{O})=\sum_{imm^{\prime }}\int_{0}^{\infty
}d\omega \overline{\rho (\omega )}_{\phi _{i}mm^{\prime }}O(\omega )_{\phi
_{i}mm^{\prime }} 
\]
or to the {\it weak limit } 
\[
W\lim_{t\rightarrow \infty }\widehat{\rho }(t)=\widehat{\rho }%
_{*}=\sum_{imm^{\prime }}\int_{0}^{\infty }d\omega \overline{\rho (\omega )}%
_{\phi _{i}mm^{\prime }}(\omega ,m,m^{\prime }|_{\phi _{i}} 
\]
where only the diagonal-singular terms remain showing that the {\it system
has decohered} in the energy.

{\bf Remarks}

i.- It looks like that decoherence takes place without a coarse-graining, or
an environment. It is not so, the van Hove choice (\ref{2}) and the mean
value (\ref{W}) are a restriction of the information as effective as the
coarse-graining is to produce decoherence.

ii.-Theoretically decoherence takes place at $t\rightarrow \infty $.
Nevertheless, for atomic interactions, the {\it characteristic decoherence
time} is $t_{D}=10^{-15}seg$\cite{DT}$.$ For macroscopic systems this time
is even smaller (e.g. $10^{-38}seg.).$ Models with two characteristic times
(decoherence and relaxation) can also be considered \cite{FP}.

\subsection{Decoherence in the other variables.}

By a change of basis we can diagonalize the $\overline{\rho (\omega )}_{\phi
_{i}mm^{\prime }}$ in $m$ and $m^{\prime }$:

\[
\rho (\omega )_{\phi _{i}mm^{\prime }}\rightarrow \rho (\omega )_{\phi
_{i}pp^{\prime }}=\rho _{\phi _{i}}(\omega )_{p}\,\delta _{pp^{\prime }}. 
\]

in a new basis orthonormal $\{|\omega ,p\rangle _{\phi _{i}}\}.$ Therefore $%
\rho _{\phi _{i}}(\omega )_{p}\,\delta _{pp^{\prime }}.$is now diagonal in
all its coordinates in a{\it \ final local pointer basis in each }$D_{\phi
_{i}}$, which, in the case of the observables is $\{$ $|\omega ,p,p^{\prime
})_{\phi _{i}}$, $|\omega ,\omega ^{\prime },p,p^{\prime })_{\phi _{i}}\}$
(i. e. essentially $\{|\omega ^{\prime },p^{\prime }\rangle _{\phi _{i}}\}),$
so in this pointer basis we have obtained a {\it boolean quantum mechanic
with no interference terms} and we have the weak limit: 
\[
W\lim_{t\rightarrow \infty }\widehat{\rho }(t)=\widehat{\rho }%
_{*}=\sum_{ip}\int_{0}^{\infty }d\omega \overline{\rho _{\phi _{i}}(\omega )}%
_{p}(\omega ,p,p|_{\phi _{i}} 
\]
or in the case of $\widehat{P}$ with continuous spectra: 
\begin{equation}
W\lim_{t\rightarrow \infty }\widehat{\rho }(t)=\widehat{\rho }%
_{*}=\sum_{i}\int_{0}^{\infty }d\omega \int_{p\epsilon D_{\phi _{i}}}dp^{N}%
\overline{\rho (\omega )_{\phi _{i}}}(\omega ,p,p|_{\phi _{i}}  \label{Y}
\end{equation}
the only case that we will consider below

\section{The classical statistical limit.}

a.- Let us now take into account the Wigner transforms. {\it There is no
problem for regular operators }which are considered in the standard theory%
{\it .} Moreover these operators are irrelevant since they disappear after
decoherence.

b.- So we must only consider the singular ones as 
\[
\widehat{O}_{S}=\sum_{i}\int_{p\epsilon D_{\phi _{i}}}dp^{N}\int_{0}^{\infty
}O_{\phi _{i}}(\omega ,p)|\omega ,p\rangle _{\phi _{i}}\langle \omega
,p|_{\phi _{i}}d\omega 
\]
where now the $\widehat{P}$ have continuous spectra. So 
\[
\widehat{O}_{S}=\sum_{i}O_{\phi _{i}}(\widehat{H},\widehat{P_{\phi _{i}})}%
=\sum_{i}\widehat{O}_{S\phi _{i}} 
\]
But $\widehat{H},\widehat{P_{\phi _{i}}}$ commute thus 
\[
symb\widehat{O}_{S}=O_{S}(\phi )=\sum_{i}O_{\phi _{i}}(H(\phi ),P_{\phi
_{i}}(\phi ))+0(\hbar ^{2}) 
\]
and if $O_{\phi _{i}}(\omega ,p)=\delta (\omega -\omega ^{\prime })\delta
(p-p^{\prime })$ we have 
\[
symb|\omega ^{\prime },p^{\prime }\rangle _{\phi _{i}}\langle \omega
^{\prime },p^{\prime }|_{\phi _{i}}=\delta (H(\phi )-\omega ^{\prime
})(P_{\phi _{i}}(\phi )-p) 
\]
({\it really up to} $0(\hbar ^{2}),$ but for the sake of simplicity we will
eliminate these symbols from now on$).$

Let us now consider the singular dual, the $symb\widehat{\rho }_{S}$ as the
functional on ${\cal M}$ that must satisfy eq. (\ref{F}) that now reads 
\[
(symb\widehat{\rho }_{S}|symb\widehat{O_{S}})=(\widehat{\rho }_{S}|\widehat{O%
}_{S}) 
\]

Then we define a density function $\rho _{S}(\phi )=symb\widehat{\rho }_{S}$ 
$=\sum_{i}\rho _{\phi _{i}S}(\phi )$ such that 
\begin{equation}
\sum_{i}\int d\phi ^{2(N+1)}\rho _{\phi _{i}S}(\phi )O_{\phi _{i}S}(\phi
)=\sum_{i}\int_{p\epsilon D_{\phi _{i}}}\int_{0}^{\infty }\rho _{\phi
i}(\omega ,p,)O_{\phi _{i}}(\omega ,p)d\omega dp^{N}  \label{L.3.10'}
\end{equation}
$\widehat{\rho _{S}},$ is constant of the motion, so $\rho _{\phi _{i}}(\phi
)=f(H(\phi ),P_{\phi _{i}}(\phi )).$ Then we {\it locally define} {\it at} $%
D_{\phi _{i}}$ the local action-angle variables ($\theta ^{0},\theta
^{1},...,\theta ^{N},J_{\phi _{i}}^{0},J_{\phi _{i}}^{1},...,J_{\phi
_{i}}^{N}),$ where $J_{\phi _{i}}^{0},$ $J_{\phi _{i}}^{1},...,J_{\phi
_{i}}^{N}$ would just be $H,P_{\phi _{i}1},...,P_{\phi _{i}N}$ and we make
the {\it canonical transformation} $\phi ^{a}\rightarrow \theta _{\phi
_{i}}^{0},\theta _{\phi _{i}}^{1},...,\theta _{\phi _{i}}^{N},$ $H,P_{\phi
_{i}1},...,P_{\phi _{i}N}$ so that 
\[
d\phi ^{2(N+1)}=dq^{(N+1)}dp^{(N+1)}=d\theta _{\phi _{i}}^{(N+1)}dHdP_{\phi
_{i}}^{N} 
\]

Now we will integrate of the functions $f(H,P_{\phi _{i}})=f(H,P_{\phi
_{i}},...,P_{\phi _{i}})$ using the new variables. 
\[
\int_{D_{\phi _{i}}}d\phi ^{2(N+1)}f(H,P_{\phi _{i}})=\int_{D_{\phi
_{i}}}d\theta _{\phi _{i}}^{(N+1)}dHdP_{\phi _{i}}^{N}f(H,P_{\phi
_{i}})=\int_{D\phi _{i}}dHdP_{\phi _{i}}^{N}C_{\phi _{i}}(H,P_{\phi
_{i}})f(H,P_{\phi _{i}}) 
\]
where we have integrated the angular variables $\theta _{\phi
_{i}}^{0},\theta _{\phi _{i}}^{1},...,\theta _{\phi _{i}}^{N}$, obtaining
the {\it configuration volume} $C_{\phi _{i}}(H,P_{\phi _{i}})$ of the
portion of the hypersurface defined by $(H=const.,P_{\phi _{i}}=const.)$ and
contained in $D_{\phi _{i}}.$ So eq. (\ref{L.3.10'}) reads 
\[
\sum_{i}\int_{p\epsilon D_{\phi _{i}}}\int_{0}^{\infty }\rho _{\phi
i}(\omega ,p,)O_{\phi _{i}}(\omega ,p)d\omega dp^{N}=\sum_{i}\int dHdP_{\phi
_{i}}^{N}C_{\phi _{i}}(H,P_{\phi _{i}})\rho _{\phi _{i}S}(H,P_{\phi
_{i}})O_{\phi _{i}S}(H,P_{\phi _{i}}) 
\]
for any $O_{\phi _{i}}(\omega ,p)$ so $\rho _{S\phi _{i}}(H,P)=\frac{1}{%
C_{\phi _{i}}}\rho _{\phi _{i}}(H,P)$ for $\phi \epsilon {\cal D}_{\phi
_{i}} $ and 
\[
\rho _{S}(\phi )=\rho _{*}(\phi )=\sum_{i}\frac{\rho _{\phi _{i}}\left(
H(\phi ),P_{\phi _{i}}(\phi )\right) }{C_{\phi _{i}}(H,P_{\phi _{i}})} 
\]
Putting $\rho _{\phi _{i}}(\omega ,p)=\delta (\omega -\omega ^{\prime
})\delta ^{N}(p-p^{\prime })$ for some $i$ and all other $\rho _{\phi
_{j}}(\omega ,p)=0$ for $j\neq i,$ we have 
\[
symb(\omega ^{\prime },p^{\prime },(\phi )|_{\phi _{i}}=\frac{\delta \left(
H(\phi )-\omega ^{\prime }\right) \delta ^{(N)}\left( P(\phi )-p_{\phi
_{i}}^{\prime }\right) }{C_{\phi _{i}}(H,P_{\phi _{i}})} 
\]

c.- Moreover the $symb$ of eq.(\ref{Y}) reads

\begin{equation}
\rho _{S}(\phi )=\rho _{*}(\phi )=\sum_{i}\int_{p\epsilon D_{\phi
_{i}}}dp\int_{0}^{\infty }d\omega \rho _{\phi _{i}}(\omega ,p)\frac{\delta
\left( H(\phi )-\omega \right) \delta ^{(N)}\left( P(\phi )-p_{\phi
_{i}}\right) }{C_{\phi _{i}}(H,P_{\phi _{i}})}  \label{3}
\end{equation}

So we have obtained a decomposition of $\rho _{*}(\phi )=$ $\rho _{S}(\phi )$
in classical hypersurfaces ($H=\omega ,$ $P_{\phi _{i}}(\phi )=p_{\phi
_{i}}),$ containing {\it chaotic trajectories} (since the system is not
integrable), summed with different weight coefficients $\rho _{\phi
_{i}}\left( \omega ,p\right) /C_{\phi _{i}}(H,P_{\phi _{i}})$.

d.- Finally only after decoherence the positive definite diagonal-singular
part remains and from eqs. (\ref{Z}$_{2}$) and (\ref{3})we see that 
\[
\rho _{\phi _{i}}(\omega ,p)\geq 0\Rightarrow \rho _{*}(\phi )\geq 0 
\]
so the {\it classical statistical limit }is obtained.

\section{The classical limit.}

The classical limit can be decomposed in the following processes

\[
Quantum\text{ }Mechanics-(\text{decohence})\rightarrow Boolean\text{ }Quantum%
\text{ }Mechanics-(\text{symb and }\hbar \rightarrow 0\text{ })\rightarrow 
\]
\[
Classical\text{ }Statistical\text{ }Mechanics-(\text{choice of a trajectory}%
)\rightarrow Classical\text{ }Mechanics 
\]
where the first two have been explained. It only remains the last one: For $%
\tau (\phi )=\theta _{\phi _{i}}^{0}(\phi )$ and at any fixed $t$ we have 
\[
\sum_{i}\int_{D_{\phi _{i}}}\delta (\tau (\phi )-\tau _{0}-\omega t)\delta
(\theta _{\phi _{i}}(\phi )-\theta _{\phi _{i}0}-p_{\phi _{i}}t)d\tau
_{0}d\theta _{\phi _{i}0}=1 
\]
then we can include this $1$ in decomposition (\ref{3}) and we obtain 
\[
\rho _{*}(\phi )=\sum_{i}\int \frac{\rho _{\phi _{i}}(\omega ,p_{\phi _{i}})%
}{C(\omega ,p_{\phi _{i}})}\delta (H(\phi )-\omega )\delta (P_{\phi
_{i}}-p_{\phi _{i}})\delta (\tau (\phi )-\tau _{0}-\omega t)\delta (\theta
_{\phi _{i}}(\phi )-\theta _{\phi _{i}0}-p_{\phi _{i}}t)d\omega d^{N}p_{\phi
_{i}}d\tau _{0}d\theta _{\phi _{i}0} 
\]
namely a sum of {\it classical chaotic trajectories} satisfying: 
\[
H(\phi )=\omega ,\text{ }\tau (\phi )=\tau _{0}+\omega t),\qquad P_{\phi
_{i}}=p_{\phi _{i}},\text{ }\theta _{\phi _{i}}(\phi )=\theta _{\phi
_{i}0}+p_{\phi _{i}}t 
\]
weighted by $\frac{\rho _{\phi _{i}}(\omega ,p_{\phi _{i}})}{C(\omega
,p_{\phi _{i}})}$ ,where we can choose any one of them$.$ In this way the
classical limit is completed, in fact we have found the classical limit of a
quantum system since we have obtained the classical trajectories, so the 
{\it correspondence principle} is also obtained as a theorem.

\section{Conclusion.}

i.- We have defined the classical limit in the non-integrable case.

ii.- Essentially we have presented a {\it minimal formalism for quantum chaos%
}\cite{QC}.

iii.- We have deduced the correspondence principle.

\end{document}